\documentclass[twocolumn,prl,aps,amsmath,
               superscriptaddress,showpacs]{revtex4}

\usepackage{epsfig}
\usepackage{float}

\def\gd         {\delta} 
\def\gee        {\epsilon}

\def\go         {\omega}
\def\qgo        {\qq,\omega}
\def\qlim       {\qq\rightarrow{\bf 0}}
\def\gr         {\rho}

\def\capo       {\right.\\ \left.}

\def\la         {\langle}
\def\ra         {\rangle}
\def\kk         {{\bf k}}
\def\qq         {{\bf q}}
\def\rr         {{\bf r}}
\def\vv         {{\bf v}}
\def\KK         {{\bf K}}
\def\GG         {{\bf G}}

\def\SS         {{\bf S}}
\def\SSt        {{\bf \tilde{S}}}

\def\WW         {{\bf W}}

\def\RR         {{\bf R}}
\def\QQ         {{\bf Q}}
\def\PP         {{\bf P}}
\def\PPt        {{\bf \tilde{P}}}

\def\ff         {{\bf f}}

\def\dPPt       {{\bf \delta \tilde{P}}}

\def\fxc        {\ff_{xc}\(\qgo\)}
\def\sfxc       {\ff_{xc}}
\def\fxcu       {\ff^{\(1\)}_{xc}\(\qgo\)}

\def\sfxcu      {\ff_{xc}^{\(1\)}}

\def\sio        {SiO$_2$\,}

\renewcommand{\[}{\left[}
\renewcommand{\]}{\right]}
\renewcommand{\(}{\left(}
\renewcommand{\)}{\right)}

\restylefloat{figure}

\begin{document}
%
\title{Bound excitons in time--dependent density--functional--theory:
       \\ optical and energy--loss spectra}
\author{Andrea Marini}
\affiliation{
Departamento de F\'\i sica de Materiales,
Facultad de Ciencias Qu\'\i micas, UPV/EHU,
Centro Mixto CSIC--UPV/EHU  and  Donostia International Physics Center.
E--20018 San Sebasti\'an, Basque Country, Spain}
\author{Rodolfo Del Sole}
\affiliation{
 Istituto Nazionale per la Fisica della Materia e Dipartimento di Fisica
 dell'Universit\`a \\ di Roma ``Tor Vergata'',
 Via della Ricerca Scientifica, I--00133 Roma, Italy}
\author{Angel Rubio}
\affiliation{
Departamento de F\'\i sica de Materiales,
Facultad de Ciencias Qu\'\i micas, UPV/EHU,
Centro Mixto CSIC--UPV/EHU  and  Donostia International Physics Center.
E--20018 San Sebasti\'an, Basque Country, Spain}
\date{\today}

\begin{abstract} 
A robust and efficient frequency dependent and non--local 
exchange--correlation $f_{xc}\(\rr,\rr';\go\)$ is derived by imposing
time--dependent density--functional theory\,(TDDFT) to reproduce the 
many--body diagrammatic expansion of the Bethe--Salpeter polarization 
function. As an illustration, we compute the optical spectra of LiF, \sio
and 
diamond and the finite momentum transfer energy--loss spectrum of LiF.
The TDDFT results reproduce extremely well the excitonic effects embodied
in the Bethe--Salpeter approach, both for strongly bound and resonant 
excitons. We provide a working expression for $f_{xc}$ that is fast to 
evaluate and easy to implement.
\end{abstract} 
\maketitle

Since the 30's, excitons have been ubiquitous in our understanding of the
optics 
of bulk materials, surfaces, nanostructures and 
organic/bio--molecules~\cite{rmp}. Only recently, however, has the first 
principle description of excitons in the optical absorption of extended 
systems been achieved, by solving the Bethe-Salpeter equation\,(BSE) 
of Many--Body Perturbation Theory\,(MBPT)~\cite{exc}. 
The solution of the BSE is usually cast into an equivalent Hamiltonian 
problem whose dimension  increases with the number of $\kk$--points and 
number of valence and conduction bands. However, even if the BSE results 
reproduce well the experimental spectra for semiconductors and insulators,
the complexity of the calculations impedes the application of this 
technique to large systems such as nanostructures and complex surfaces.

An alternative approach to the study of correlation in many--body systems 
is given by density--functional theory, in its static\,(DFT)~\cite{nobel} 
and time dependent formulations\,(TDDFT)~\cite{tddft}.
Similar to the paradigm of DFT for ground-state properties, TDDFT has 
emerged as a very powerful tool for the description of excited states. In 
principle TDDFT is exact for neutral excited--state properties,
and its simplicity relies on the fact that two-point response functions 
are needed instead of the four-point function of the BSE~\cite{rmp}.
TDDFT casts all many-body effects into the dynamical exchange-correlation 
kernel $f_{xc}\(\rr,t,\rr';t'\)=\gd v_{xc}\(\rr,t\)/\gd\gr\(\rr',t'\)$,
where $v_{xc}\(\rr,t\)$ is the corresponding time--dependent 
exchange--correlation potential.
It was early recognized~\cite{perdew} that, in extended systems, the 
standard approximations for $v_{xc}$ --\,local density\,(LDA) or 
generalized gradient\,(GGA)\,-- that work extremely well for ground state 
properties, fail to describe,  among other effects, the band--gap of 
insulators and semiconductors and the excitonic effects in the optical and 
energy--loss spectra~\cite{rmp}.
Recently, promising results have been obtained within a polarization 
dependent functional derived in the framework of current-DFT~\cite{boeij}
and within the exact-exchange DFT approach~\cite{fexx}. The calculated 
optical spectrum of silicon exhibits excitonic effects in 
qualitative agreement with experiment. However, empirical cutoffs were 
introduced to construct $f_{xc}$~\cite{xcpar}, that somehow account for 
the screening of the electron--hole interaction.
Furthermore, to--date the calculations of the absorption spectra of solids 
beyond time--dependent\,LDA were performed in semiconductors characterized 
by weak continuum excitonic 
effects~\cite{boeij,fexx,lucia,rod,gianni,francesco}. 
Consequently it remains open whether or not strong electron--hole 
effects\,(e.g. bound excitons) in the optical and energy--loss spectra can 
be described within TDDFT.
The present letter resolves unambiguously this controversy,
by deriving a novel $f_{xc}$ that indeed accounts
for excitonic effects in semiconductors and wide--gap insulators.

The challenge is how to construct $f_{xc}$. To address this goal we 
benefit from the good performance of MBPT response functions and build an 
$f_{xc}$ that mimics those results. This $f_{xc}$ is derived by imposing 
TDDFT to reproduce the perturbative expansion of the BSE in terms of the 
screened Coulomb interaction at any order. Our derivation starts from the 
usual TDDFT equation for the irreducible response function 
$\PPt\(\qgo\)$~\cite{tddft}
\begin{align}
\PPt\(\qgo\)=\PP^{\(0\)}\(\qgo\)+
 \PP^{\(0\)}\(\qgo\)\ff_{xc}\(\qgo\)\PPt\(\qgo\),
\label{irrtddft}
\end{align}
where $\PP^{\(0\)}\(\qgo\)$ is the ``exact" Kohn--Sham DFT response 
function for momentum transfer $\qq$. All quantities are two--point 
functions (matrices in reciprocal space). The microscopic dielectric 
matrix is obtained from $\PPt$ and the Coulomb potential $\vv$ as
$\varepsilon\(\qgo\)={\bf 1}-\vv\PPt\(\qgo\)$.
Now we make the connection with MBPT. First, $\PP^{\(0\)}\(\qgo\)$
is approximated by the independent--quasiparticle\,(QP) response, 
calculated in the GW scheme~\cite{hedin}. Second, we {\it assume} that 
there exists an $\fxc$ that reproduces the BSE spectra~\cite{gianni},
i.e., we impose $\PPt\(\qgo\)\equiv\PPt_{BSE}\(\qgo\)$. Without loss of 
generality, we restrict our derivation to the resonant part of 
$\PPt\(\qgo\)$:
\begin{multline}
 \tilde{P}_{\GG_1,\GG_2}\(\qgo\)=\frac{2i}{\Omega}\\
  \sum_{\KK_1,\KK_2}\Phi^{*}_{\KK_1}\(\qq,\GG_1\) 
  \tilde{S}_{\KK_1,\KK_2}\(\qgo\) \Phi_{\KK_2}\(\qq,\GG_2\),
\label{pbse}
\end{multline}
where $\Omega$ the crystal volume, and $\KK=\(c\,v\,\kk\)$ a generalized 
index to describe the space of electron--hole states. The oscillators 
${\bf \Phi}_{\KK}$ are given by $\Phi_{\KK}\(\qq,\GG\)=\la 
c\kk|e^{i\(\qq+\GG\)\cdot\rr}|v\kk-\qq\ra$, in terms of the conduction and 
valence Kohn--Sham states. In Eq.\,(\ref{pbse}) $\SSt\(\qgo\)$ is the 
solution of the BSE~\cite{rmp}
\begin{align}
\SSt\(\qgo\)=\SS^{\(0\)}\(\qgo\)+
 \SS^{\(0\)}\(\qgo\)\WW\(\qq\)\SSt\(\qgo\),
\label{bse}
\end{align}
with $S_{\KK_1\KK_2}^{\(0\)}\(\qgo\)=
      i\gd_{\KK_1\KK_2}\(\go-E_{\KK_1}^{\(\qq\)}+i0^+\)^{-1}$,
and $E_{\KK}^{\(\qq\)}=\gee^{QP}_{c_1\kk_1}-\gee^{QP}_{v_1\kk_1-\qq}$ in 
terms of the electron--hole QP energies\,($\gee^{QP}$). 
$\WW\(\qq\)$ is the Coulombic part of the Bethe--Salpeter kernel, 
$W_{\KK_1\KK_2}\(\qq\)=i\la c_1\kk_1,v_2\kk_2-\qq|W\(\rr_1,\rr_2\)
 |c_2\kk_2,v_1\kk_1-\qq\ra$,
with $W\(\rr_1,\rr_2\)$ the statically screened electron--hole 
interaction. Eq.\,(\ref{irrtddft}) can be transformed into an equation for 
$\fxc$ by putting $\PPt\(\qgo\)=\PP^{\(0\)}\(\qgo\)+\dPPt\(\qgo\)$
\begin{multline}
\PP^{\(0\)}\(\qgo\)\ff_{xc}\(\qgo\)\PP^{\(0\)}\(\qgo\)=\\
\dPPt\(\qgo\)-\dPPt\(\qgo\)\ff_{xc}\(\qgo\)\PP^{\(0\)}\(\qgo\).
\label{eqfxc}
\end{multline}
The advantage of Eq.\,(\ref{eqfxc}) is that, by expanding $\dPPt\(\qgo\)$
in terms of $W$, $\dPPt\(\qgo\)=\sum_n\dPPt^{\(n\)}\(\qgo\)$, it is 
possible to write  the n--th order ($\ff^{\(n\)}_{xc}\(\qgo\)$) 
contribution to $\fxc$ in an iterative form:
\begin{multline}
\ff^{\(n\)}_{xc}\(\qgo\)=
 \[\PP^{\(0\)}\(\qgo\)\]^{-1}
 \[
  \dPPt^{\(n\)}\(\qgo\)\(\PP^{\(0\)}\(\qgo\)\)^{-1}\capo
   -\sum_{m=1,n-1}\(-1\)^m\dPPt^{\(m\)}\(\qgo\)\ff^{\(n-m\)}_{xc}\(\qgo\)
 \],
\label{expansion}
\end{multline}
with $\dPPt^{\(0\)}\(\qgo\)=0$. This ends our derivation.

Other attempts to build $\ff_{xc}$ from MBPT rely on either fully solving 
the BSE~\cite{rod} or by imposing $\ff_{xc}$ to be static and linear in 
$W$~\cite{lucia}. However, the static constraint has to be released 
in the practical solution for the optical spectra of silicon and 
SiC~\cite{francesco}. This internal inconsistency is solved in the 
present work as Eq.\,(\ref{expansion}) provides a systematic and 
consistent treatment of the frequency dependence and local field 
effects\,(LFE) of $\fxc$ at the cost of making $\fxc$ a  non--linear 
functional of $W$. The scheme proposed in 
Refs.~\cite{lucia,gianni,francesco} appears naturally as an approximate 
solution of Eq.\,(\ref{expansion}).

We illustrate the reliability of the present TDDFT approach in three 
prototype systems: LiF, \sio and diamond~\cite{details}.
In these three systems the role of 
excitonic effects in the optical spectrum and EELS has been already 
analyzed within the BSE~\cite{chang,bslifc}.
\sio  is characterized by four strong excitonic peaks at $10.3$, $11.3$, 
$13.5$ and $17.5$ eV, none of them below the QP gap of 10.1\,eV, except 
for a bound triplet exciton optically inactive. Moreover, the exciton at 
$10.3$\,eV corresponds to a strongly  correlated resonant state with a 
large degree of spatial localization (2--3 bond lengths)~\cite{chang}.
The spectrum of LiF is dominated by a strongly bound 
exciton\,($\sim$\,3\,eV binding energy)\cite{bslifc}. Last, in diamond, 
the electron--hole interaction produces a drastic modification of the 
independent QP spectrum by shifting optical oscillator strength from high 
to low energies.
\begin{figure}[H]
\begin{center}
\epsfig{figure=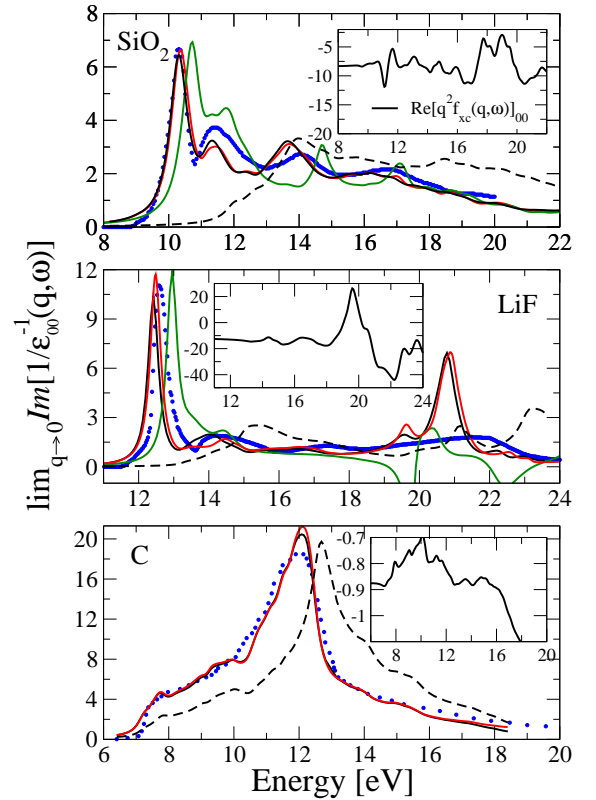,clip=,bbllx=5,bblly=30,bburx=520,
                             bbury=770,width=7.5cm}
\end{center}
\vspace{-.5cm}
\caption{
\footnotesize{Optical absorption spectra calculated within the BSE\,(black 
line), and TDDFT\,(red line) compared with experiments\,(blue 
circles)~\cite{exp} and with the independent--QP calculation\,(dashed 
line). The results obtained using a scalar $\sfxc$\,(green line) are shown 
to stress the importance of the LFE in $\sfxc$. The frequency dependence 
of the head of $\sfxc$, 
$\lim_{\qq\rightarrow{\bf 0}}\Re\[q^2\fxc\]_{{\bf 0},{\bf 0}}$ 
is shown in the insets (note the different scale for \sio, LiF and C).
}}
\label{f1}
\end{figure}
The TDDFT results for the optical absorption of those three elements using 
the first order $\ff_{xc}^{\(1\)}$\ of Eq.\,(\ref{expansion})
($\ff_{xc}^{\(1\)}=\[\PP^{\(0\)}\]^{-1}\dPPt^{\(1\)}\[\PP^{\(0\)}\]^{-1}$)
are compared in Fig.\,\ref{f1}\,(red--line) with the  BSE spectra\,(black 
line) and experiments\,(blue circles)~\cite{exp}.
To put in evidence the strong role played by excitonic effects, the 
independent--QP spectrum is also shown\,(dashed line). In all three cases 
the agreement between TDDFT and BSE is excellent. This good performance of 
$\ff^{\(1\)}_{xc}$ holds not only for the head of $\varepsilon\(\qgo\)$ 
but also for the off--diagonal elements. 

The computational cost of our scheme is mainly given by the size of 
$\varepsilon\(\qgo\)$ that is proportional to the degree of inhomogeneity 
of the induced density, i.e. it is related to the microscopic LFE. In the 
present TDDFT calculations of the response functions we have two 
contributions to LFE: the standard contribution from the Hartree potential 
as included in a calculation within the random--phase approximation\,(RPA) 
and the one coming from the exchange-correlation effects embodied in 
$\ff_{xc}$ (e.g., electron-hole attraction). 
In all the systems we have looked at, it turns out that the size of the
LFE from $\ff_{xc}$ is smaller or equal to that of the Hartree contribution. 
This general behavior of $\ff_{xc}$ in extended systems is due to the 
spatial localization of the excitonic wavefunctions, that exceeds the 
typical length scale for the density variations\,(bond--length). 
As the proper description of LFE is achieved with matrix sizes much smaller 
than the typical size of the BSE Hamiltonian (of the order of 
$10^4\times\,10^4$), the present TDDFT approach should be computationally 
favorable with respect to BSE. 

To elucidate the role of the LFE in $\ff_{xc}$ we show in Fig.\,\ref{f1}
with 
green-line the results imposing a first--order scalar, but frequency 
dependent $f_{xc}$: $\[\ff^{\(1\)}_{xc}\(\qlim,\go\)\]_{\GG=\GG'={\bf 0}}$
(whose real part is shown in the insets of Fig.\,\ref{f1}). For diamond, 
as well as other semiconductors not shown here, this scalar $f^{(1)}_{xc}$ 
works very well\,(see Fig.\,\ref{f1}). In particular, the head of $\sfxc$ 
is a smooth function of frequency below the QP-gap. This result 
supports the long-range model of Ref.\cite{lucia}. However, \sio and LiF 
are not at all well described by this scalar $\sfxcu$.
Only the main peak is qualitatively reproduced but not at the correct 
energy, and even unphysical regions of negative absorption appear for LiF 
at high energies.
A good spectrum is only achieved when the $\sfxcu$ matrix dimension is set 
to 267$\times$\,267 and 59$\times$\,59 for \sio and LiF, respectively. 
Increasing the $\ff_{xc}$ matrix size does not introduce any change in the 
spectrum. Furthermore, the head of $\sfxcu$ is strongly frequency 
dependent in order to describe the high-energy features of the spectra 
(see inset in Fig.\,\ref{f1}). In conclusion, {\it for the description of 
the absorption and energy--loss spectra in systems with continuum 
excitonic effects (e.g. diamond) the  frequency dependence and the 
microscopic LFE of $\ff_{xc}$ are not important while they become crucial
in wide--gap insulators with bound--excitons\,(e.g. LiF) or strongly 
correlated resonant states\,(e.g. \sio)}.

\begin{figure}[H]
\begin{center}
\epsfig{figure=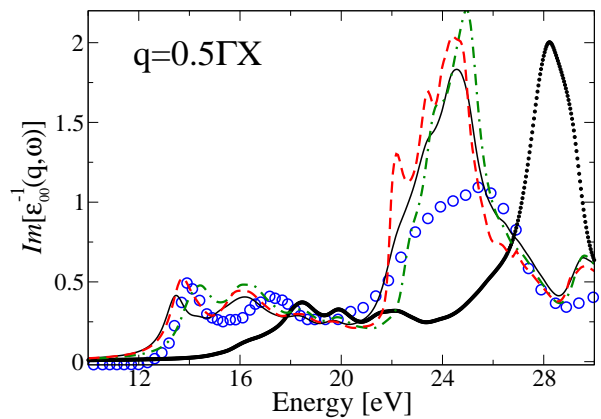,clip=,bbllx=88,bblly=20,
               bburx=590,bbury=745,height=8cm,angle=-90}
\end{center}
\vspace{-.5cm}
\caption{
\footnotesize{ (Color online) Calculated EELS of LiF for a momentum 
transfer of $\qq=0.5\Gamma\,X$: BSE\,(full line), 1$^{st}$ order 
$\ff_{xc}$\,(dot--dashed line), 2$^{nd}$ order $\ff_{xc}$\,(dashed line),
and independent-QP\,(dots). Experiment\,(circles) are taken from 
Ref.~\cite{shirley}.}}
\label{f2}
\end{figure}

To show the robustness and transferability of the perturbative approach to 
$\ff_{xc}$ we calculated the EELS of LiF for a {\it finite transfer 
momentum} $\qq$ along the $\Gamma X$ direction where previous BSE 
calculations and experimental results are available~\cite{shirley}.
This is a stringent test as  the description of EELS needs causal response 
functions, including the anti--resonant part~\cite{eelsnote}.
In the present theory a causal $\ff_{xc}$ can be easily obtained by 
inserting in  Eq.\,(\ref{expansion}) the causal  $\PPt\(\qgo\)$ obtained 
from the time--ordered $\PPt$ of MBPT. The results of this calculation are 
presented in Fig.\,\ref{f2} for a 1$^{st}$ order\,(dot--dashed line) and a 
2$^{nd}$ order\,(dashed line) causal $\ff_{xc}$.
In contrast to the results of Fig.\,\ref{f1} for the optical absorption,
we need to go to a second order causal $\ff_{xc}$ to restore the good
agreement between TDDFT and BSE. Still, a first order $\fxcu$ gives very 
reasonable EEL spectrum. The results for the optical absorption and EEL 
{\it highlight that the first order $\fxcu$ embodies all relevant many body 
effects at the BSE level even though it is a contracted two--point
function}.
Moreover higher order contributions to $\ff_{xc}$ yield minor 
modifications to the spectrum because strong cancellations occur at 
any order, except for the first, of the perturbative expansion  of $\fxc$ 
[Eq.\,(\ref{expansion})].
A numerical remark about the results in Figs.\,\ref{f1}--\ref{f2} is
relevant now. The direct application of Eq.\,(\ref{expansion}) for 
$\ff_{xc}$ leads to spurious oscillations in the calculated optical 
spectra. Those oscillations are moderate in the case of diamond but they 
tend to destroy the spectra for the case of LiF and \sio. The
intensity of the oscillations increases with the order of 
$\ff_{xc}^{\(n\)}$ and eventually, gives rise to non--physical regions of 
negative absorption.
The reason for this numerical pathology  stems from the way the 
Bethe--Salpeter kernel acts on the  spectra: (i)\,redistributing the 
optical oscillator strength, (ii)\,shifting rigidly the spectra to 
account for the diagonal of the Bethe--Salpeter kernel 
$\Delta_{\qq}=W_{\KK,\KK}\(\qq\)$, that should vanish in the limit of 
infinite $\kk$--point sampling.
However with a finite $\kk$--point grid, corresponding to a fully 
converged BSE spectrum, we get $\Delta_{\qlim}\sim$\,$-0.3$\,eV in 
diamond, and $\Delta_{\qlim}\sim$\,$-0.9$\,eV in LiF and \sio.
The diagonal of $\WW\(\qq\)$ appears in $\fxc$ through a series expansion
in $\Delta_{\qq}$~\cite{deltanote} that  is meaningful only when 
$\Delta_{\qq}$ is sufficiently small. As this is not the case for \sio and 
LiF, natural oscillations are found in the {\it na\"{\i}ve} application of 
the perturbative expansion of Eq.\,(\ref{expansion}). To circumvent this 
issue  we included the diagonal part of $\WW\(\qq\)$ into the independent 
QP response function $\PP^{\(0\)}\(\qgo\)$ and let $\fxcu$ account 
explicitly for the off--diagonal contributions to the Bethe--Salpeter 
kernel. Using this idea the higher order corrections to $\fxcu$ are not 
only well defined but numerically stable at all orders with the same 
$\kk$--point sampling used in a standard BSE calculation. 

In conclusion, we have shown that TDDFT with a first order $\fxcu$ 
reproduces the optical and energy--loss spectra for a large class of 
materials: insulators and wide--gap insulators. In particular, bound 
excitons are described within TDDFT. Still, the direct implementation of 
$\fxcu$ is cumbersome. However, by looking at the analytic properties of 
$\dPPt^{\(1\)}\(\qgo\)$\,(for an explicit expression see 
Ref.~\cite{gianni}) we can single out the contribution of degenerate 
non--interacting electron--hole states in Eq.\,(\ref{pbse})  and write a 
general expression for $\fxcu$:
\begin{multline}
 \fxcu=\frac{2}{\Omega} \[\PP^{\(0\)}\(\qgo'\)\]^{-1} 
 \sum_{\KK}\[\frac{\RR^{\(\qq\)}_{\KK}+\RR_{\KK}^{\(\qq\)\dagger}}
                  {\go'-E^{\(\qq\)}_{\KK}+i0^+}\capo
  +\frac{\QQ^{\(\qq\)}_{\KK}}{\(\go'-E^{\(\qq\)}_{\KK}+i0^+\)^2}\]
 \[\PP^{\(0\)}\(\qgo'\)\]^{-1}.
\label{eqsimpl}
\end{multline}
Here $\go'=\go+\Delta_{\qq}$ and the sum runs to all independent 
electron--hole states with residual
$\[R^{\(\qq\)}_{\KK}\]_{\GG_1,\GG_2}= 
 \sum_{\KK', E^{\(\qq\)}_{\KK'}\neq E^{\(\qq\)}_{\KK}} 
 \frac{\Phi^{*}_{\KK}\(\qq,\GG_1\)W_{\KK,\KK'}\(\qq\) 
 \Phi_{\KK'}\(\qq,\GG_2\)}{E^{\(\qq\)}_{\KK}-E^{\(\qq\)}_{\KK'}}$
for non--degenerate states, and
$\[Q^{\(\qq\)}_{\KK}\]_{\GG_1,\GG_2}=
 \sum_{\KK', E^{\(\qq\)}_{\KK'}=E^{\(\qq\)}_{\KK}}
 \Phi^{*}_{\KK}\(\qq,\GG_1\)W_{\KK,\KK'}\(\qq\) \Phi_{\KK'}\(\qq,\GG_2\)$
for degenerate states. Eq.\,(\ref{eqsimpl}) is the main practical result 
of this letter. It is very fast to compute~\cite{speed} as it has  the 
form of a  non--interacting polarization function with modified 
residuals\,($Q,R$) that are evaluated only once as a result of two simple 
matrix--vector multiplications.
Also, Eq.\,(\ref{eqsimpl}) can be made causal and be extended to higher 
orders of the perturbative expansion of $\fxc$. This implementation opens 
the way for calculations of  the response function of nanostructures and 
low--dimensional systems within TDDFT. Work along this line is under 
progress.

This work has been supported by the NANOPHASE Research Training Network
(HPRN-CT-2000-00167), Spanish MCyT\,(MAT 2001--0946), INFM PAIS CELEX and 
MIUR Cofin 2002. We thank G.\,Adragna, L.\,Reining, V.\,Olevano
and F.\,Sottile for enlightening discussions and for providing us 
with\,Ref.\cite{francesco} before publication.
We also thank L. Wirtz and C. Hogan for a critical reading.



\begin{thebibliography}{100}
\bibitem{xcpar}		In Ref.~\cite{fexx} a cutoff was introduced
 to avoid the collapse of the optical spectra, and
 in Ref.~\cite{boeij} to fit the experimental E$_1$--E$_2$ peak intensities 
 and to rigidly shift the spectrum.
\bibitem{eelsnote}
 This crucial point has been overlooked in the previous derivations of 
 $\ff_{xc}$ based on MBPT~\cite{lucia,francesco}, as they looked only at 
 the absorption spectra.
\bibitem{deltanote}
 The contribution of $\Delta_{\qq}$ to $\fxc$, 
 $\ff_{xc}^{\Delta}\(\qgo\)$, can be isolated from Eq.\,(\ref{expansion}).
 $\ff_{xc}^{\Delta}\(\qgo\)=-\sum_n \Delta_{\qq}^n 
 \partial^n\[\PP^{\(0\)}\(\qgo\)\]^{-1}/\partial\go^n$. This series, in 
 general, oscillates and cannot be treated pertubatively. The same holds 
 for the description of QP--shifts.
\bibitem{speed} In the case of diamond the solution of the TDDFT equation
 using Eq.\,(\ref{eqsimpl}) is two orders of magnitude faster than the
 direct diagonalization of the BSE hamiltonian.
\end{thebibliography}
\end{document}